*Analytical Article*

# The Conundrum of the Pension System in India: A Comprehensive Study in the Context of India's Growth Story

[1]**Deeti Aditya**
[1](MBA), NET-JRF, PGRRCDE, Osmania University, Telangana, India.

***Abstract:*** *India is the largest democracy in the world and has recently surpassed China to be the highest-populated country, with an estimated 1.425 billion (approximately 18% of the world population). Moreover, India's elderly population is projected to increase to 138 million by 2035. Indian economy is already reeling under the pressure of exorbitant pension liabilities of the government for existing pensioners. As such, India has introduced a National Pension System (NPS), which is a Defined Contribution Scheme for employees joining government service on or after 1$^{st}$ January 2004, bidding adieu to the age-old, tried and tested Old Pension System (OPS) which is a Direct Benefit Scheme, in vogue in India since the British Raj. This is an epoch-making move by the government as it seeks to inculcate Disciplined Saving among the people while significantly reducing the government burden by reducing the Pension Liabilities of the Central and State Governments. This paper aims to analyse various features and intricacies of the NPS and address the claims of various stakeholders like the Central Government, State Government, Employees, Pensioners, etc. In light of the above, and taking cognisance of the fact that many states such as Rajasthan, Chattisgarh, Jharkhand, etc, have exited the NPS scheme and have sought back their share of NPS employee and employer contribution, this study is relevant to address the current economic and fiscal issues of India to propel towards the ambitious goal of becoming a $ 5 trillion dollar economy by 2025.*

## I. INTRODUCTION

The Old Pension System (OPS) has been serving the needs and aspirations of the citizens since independence, and it stood the test of time. But, in light of population growth in India and the fact that the demography is gradually shifting towards an older population, the OPS seems unsustainable. This is due to the difference between the growth rate of revenue and the growth rate of Pension Liabilities of the government. This results in the government earmarking more and more revenue proceeds towards Pension payments, which is unsustainable in the long run.

The two Pension schemes are fundamentally different in the sense that the OPS is a "Direct Benefit Scheme" that also accounts for the cost of living through regular "Dearness Relief (DR)" to the pensioners, whereas the National Pension System (NPS) is a "Defined Contribution Scheme" that pools the employee and employer contributions of 10% and 14% respectively of the sum of the Basic Pay and Dearness Allowance (DA).

Moreover, the development of the Health sector since independence has resulted in a decline in Infant Mortality Rate (IMR), Maternal Mortality Rate (MMR), Total Fertility Rate (TFR), etc, by a huge margin while also increasing the Life Expectancy to 67.2 years which is further bound to increase. All these factors indicate a further burden on the exchequer for pension payments.

Apart from the above, there is also a need to address the income distance among people above 65 years, which is fairly contrasting in the case of Organised and Unorganised workers as their pension amount varies by a huge margin. While organised workers are served under NPS, the unorganised sector workers are served with several alternative pension schemes like Atal Pension Yojana (APY), Shram-Yogi Maan Dhan Yojana, etc, in which the maximum monthly pension is capped at ₹ 3,000/- or ₹ 5,000/-. This is much below the NPS pension and thus inconsistent with establishing a welfare state, as mentioned in the DPSPs.

It is clear that the NPS is the key to addressing all these problems, but it has its fair share of issues, such as not guaranteeing a defined pension because it is a market-linked fund management scheme whose returns are subject to market volatility. Moreover, NPS is still a voluntary scheme for Private companies and corporations, meaning that most private sector employees have not been brought under the Pension net. All these hinder the true potential of India's Social security sector, and the recent protests of many Indian states like Chattisgarh, Rajasthan, etc, to withdraw from NPS bolster the cause to pursue a comprehensive study on this topic.

Though the NPS has multiple issues to be addressed, this paper is limited to analysing the claims of employees that the NPS does not provide guaranteed pension amounts.

## II. LITERATURE REVIEW

The NPS is a defined contribution scheme as opposed to the OPS, which is a Direct Benefit Scheme. Hence, the subscriber of an NPS account will have to contribute 10% of his/her Basic Pay plus [1]Dearness Allowance (DA) as a defined monthly contribution throughout their service. This would be complemented by the employer contribution of 14% of Basic Pay plus Dearness Allowance (DA). The NPS Trust, incorporated by the Pension Fund Regulator PFRDA has launched an NPS Pension calculator which gives the value of Maturity proceeds and the Annuity value post superannuation by considering various factors such as subscriber's age, the sum of employee and employer contribution, the lumpsum amount needed, the share of annuity in the total corpus, expected return on investment, the required rate of annuity etc.

In the case of OPS, the pension is approximated as 50% of the last drawn salary at the time of superannuation. Moreover, the employees under OPS are eligible for retirement gratuity of a maximum ₹ 20 lakh, subject to the length of the employee's service. This feature is not present in NPS and hence has to be offset by withdrawing some lumpsum from the NPS retirement corpus itself. This is one reason of contention for the employees demanding a return to the OPS. Let us look at both these schemes objectively by considering the case of a central government employee appointed in the central pay scale of Level-10 (GP 5400) with an initial Basic Pay @ ₹ 56,100/- and DA @ 42%. In the case of OPS, we make the following assumptions for calculating pension:

1) The age of the employee at the initial appointment is 25 years.
2) The age of superannuation is 60 years.
3) The annual increment is @ 3% of Basic Pay.
4) The average annual DA increment is @ 6% p.a.
5) The tenure of service is 35 years.
6) Present Gross salary is ₹ 1,10,000/- p.m

To find the pension for this person at 60 years of age, we employ the concept of [2]Time Value of Money and compound the present gross salary of ₹ 1,10,000/- at the rate(r) of **9% p.a.** (3% increment + 6% DA increase)

$$\textit{Future Value of Salary} = \textit{Present Value} \times \left(1 + \frac{r}{100}\right)^{\textit{Tenure of service in years}}$$

*Future Value of Salary = ₹ 1,10,000 × 1.09³⁵ = **₹ 22,45,536/- p.m.***

Therefore, the OPS turns out to be around **₹ 11,22,768/- p.m**. since it is 50% of the Last drawn salary.

*A) NPS Pension calculation*

Now, let us move to the NPS pension calculation. This can be done by using the NPS Pension calculator developed by NPS Trust by keying down the required values of age, retirement age, contribution, choice of lumpsum, share of annuity in the total corpus, expected rate of return on investment, desired rate of annuity etc. We consider the following assumptions in calculating the NPS amount:

1) The age of the employee at the initial appointment is 25 years.
2) The age of retirement is 60 years.
3) The initial Basic pay is ₹ 56,100/- p.m., and DA is @ 42%.
4) The tenure of contribution is till 60 years.
5) The expected rate of return on investment is 9%.
6) The share of the annuity in the total corpus is 75%.
7) The desired rate of annuity is 8%

The value of the total retirement corpus amounts to ₹ 5,33,49,262/- whereas the lumpsum @ 25% of the corpus amounts to ₹ 1,33,37,315/- and the remaining ₹ 4,00,11,947/- comprises the retirement annuity. Furthermore, the monthly pension trickles down to **₹ 2,66,746/- p.m.**

Now, let us analyse the change in Pension amount with the change in share of annuity in total corpus, investment portfolio opted by the fund managers under the Central Government Scheme of NPS. NPS offers four major categories of investment options: Equity (E), Corporate Debt (C), Government Securities (G) and Alternative Investment Funds (A), which have varying risk-return trends. Alternative Investment Funds (A) are highly risky, followed by Equity (E), Corporate Debt (C) and Government Securities (G). The prescribed cap on investment in various categories is as follows:

| Type of Security | Investment cap (%) |
|---|---|
| Government Securities | Upto 55% |
| Corporate Debt | Upto 45% |
| Equity | Upto 15% |
| Short-term debt | Upto 10% |
| Alternate Investments & Miscellaneous | Upto 5% |

Moreover, depending on the individual's risk appetite, three Life cycle investment options, namely, Aggressive, Moderate and Conservative Life Cycle funds with varying Equity investment caps @ 75%, 50% and 25%, are provided. Opting for these options will alter the return on investment of the employee.

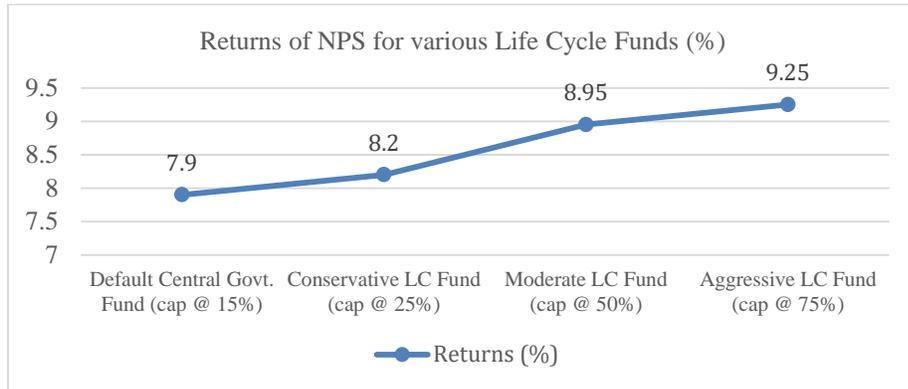

**Fig 1: Change in NPS returns with change in investment option**

$$Weighted\ Average\ Rate\ of\ Return = \frac{\sum(weight) \times (expected\ rate\ of\ return\ of\ asset\ class)}{\sum(weights)}$$

Expected return of Equity (E) = 10%
Expected return of Government Securities (G) = 7%
Expected return of Corporate Bond (C) = 8%

*The investment is first allocated to the Equity Segment, followed by Corporate Debt and Bonds and finally to Government Securities in all cases.

Similarly, now we will look into the pension amounts for various levels of Annuity share in the total corpus, i.e., 40%, 50%, 60%, 70%, 75% and 80%. The pension is calculated by considering 10% employee contribution and 14% employer contribution on the basic pay of ₹ 56,100/- and DA @ 42% of basic pay.

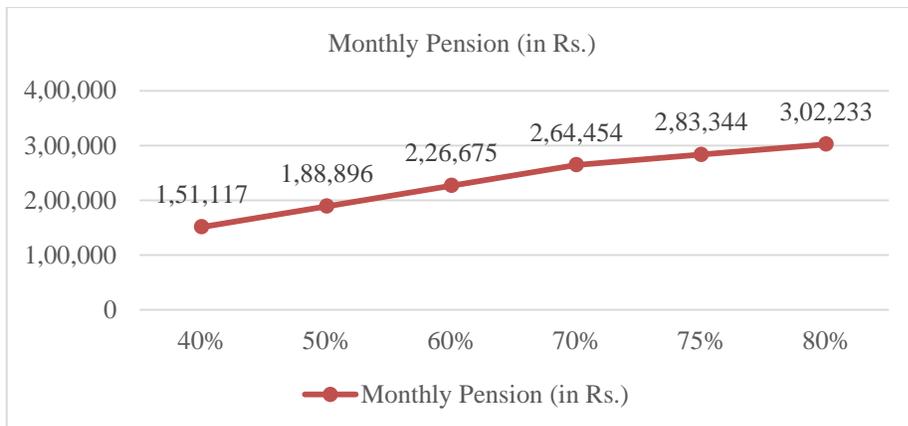

**Fig 2: Change in Monthly Pension (in Rs.) with change in share of annuity in total corpus.**

Furthermore, let us explore the possibility of increasing the employer contribution from 14% to 20% of Basic pay plus Dearness Allowance.

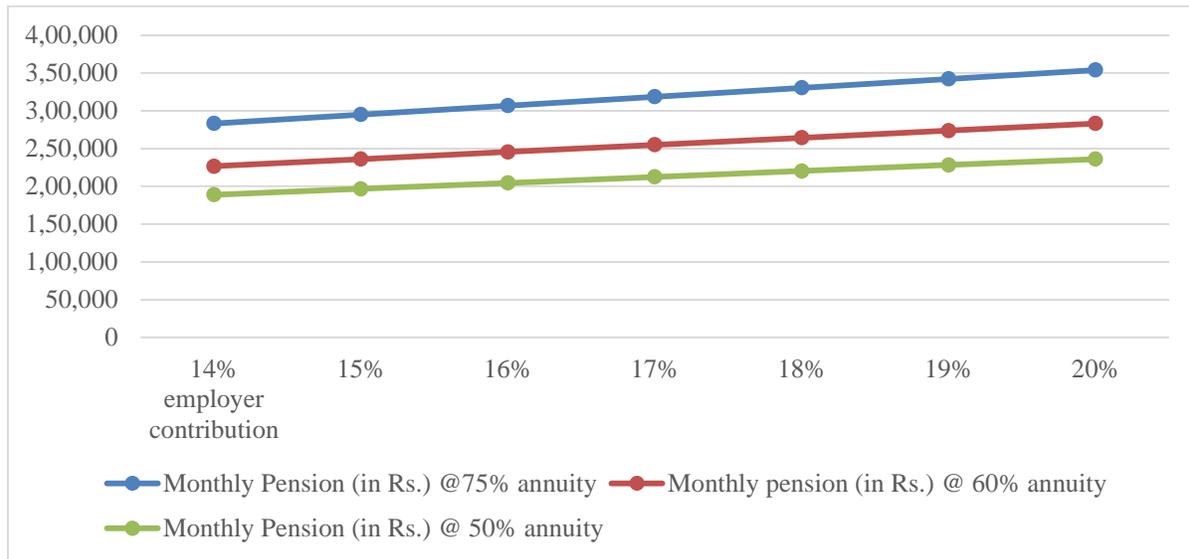

**Fig 3: Change in Monthly Pension with increase in Employer Contribution for various Annuity share options.**

### III. RESULTS AND DISCUSSION

- There is a huge difference between the Pension amount under OPS and the pension under the NPS scheme for the same employee with similar values of all remaining values and assumptions. The OPS Pension came out to be ₹ 11,22,768/- p.m., whereas the NPS Pension stood at ₹ 2,66,746/- p.m.
- It is evident from Fig 1 that the pension fund's returns increased considerably as the risk appetite of the subscriber increased. Hence, the pension amount increased with the increase in investment cap in the Equity (E) segment from 15% in the default fund to 75% in the Aggressive Life Cycle Fund.
- Moreover, from Fig 2, the Monthly pension increased significantly from ₹ 1,51,117/- to ₹ 3,02,233/- p.m. as the share of annuity increased as a percentage of Total Retirement corpus from 40% to 80%.
- From Fig 3, we find that the pension increased significantly with an increase in both the share of annuity and the employer contribution from 40% to 80% and 14% to 20%, respectively.

### IV. CONCLUSION

- The NPS monthly pension can be guaranteed for a defined amount to the tune of 15% of the last drawn pay of the employee, as in the case of OPS, where it is fixed at 50%. This is evident from Fig 3, where we can observe that by carefully juggling between the Share of Annuity and the Employer contribution, the pension can be significantly improved to ₹ 3,18,732/- p.m., which is about 14.19% of the employee's last drawn salary of ₹ 22,45,536/- p.m. with Annuity share @ 75% of the corpus and employer contribution @ 17% of Basic Pay and DA. This is also favourable to the government because of very little additional burden on the government but a significant rise in pensions.
- The NPS scheme must be tweaked a little to increase the cap on investments in Equity (E) to at least 50% even under "Auto Choice" mode, as most people do not have the required Financial Literacy and knowledge to invest by themselves through "Active Choice" mode for higher pension.
- The withdrawal of corpus must be limited to 25% of the total retirement corpus so that most of it is reinvested for an annuity, which results in a much higher monthly pension.
- The subscribers must be encouraged to invest in the [3]Tier II NPS Account to meet their needs without disturbing their Retirement corpus in the [4]NPS Tier I Account.

**Appendix 1**

Dearness Allowance (DA) is the compensatory payment component of salary to offset inflation and a rising cost of living.

**Appendix 2**

"Time Value of Money" is a concept that states that "A certain sum of money is worth more now than it is at a future time". It is basically the idea of compounding the money to get the Future value at a time in future.

**Appendix 3**

NPS Tier II Account is a type of investment account offered to the existing NPS subscribers having an NPS Tier I Account. Tier I Account is a Retirement corpus account with Tax benefits, while Tier II Account is a savings and investment account without any Tax Benefits.

**Appendix 4**

NPS Tier I Account is the Retirement corpus Account where the contributions of both the employee and the employer are pooled at the rate of 10% and 14%, respectively, to act as a payout source of monthly pension to the subscriber post superannuation, usually at 60 years.